\documentclass[12pt]{article}
\usepackage{epsfig,amsfonts,amssymb}
\usepackage{hyperref}
\usepackage{cite}
\input epsf.sty
\usepackage{cite}

\def\ZZZ{{\hbox{ Z\kern-1.6mm Z}}}
\def\RRR{{\hbox{ R\kern-2.4mm R}}}
\def\CCC{{\hbox{ C\kern-2.0mm C}}}
\def\zzz{{\hbox{z\kern-1mm z}}}

\newcommand{\qeq}{{\hbox{=\kern-2.3mm ? \kern.5mm }}}
\renewcommand{\qeq}{=}

\newcommand{\FF}{{\cal F}}

\newcommand{\OO}{{\cal O}}

\newcommand{\wt}{\widetilde}
\newcommand{\wh}{\widehat}

\newcommand{\NN}{{\cal N}}

\newcommand{\be}{\begin{equation}}
\newcommand{\ee}{\end{equation}}
\newcommand{\ben}{\begin{eqnarray}\displaystyle}
\newcommand{\een}{\end{eqnarray}}

\newcommand{\bea}[1]{\begin{eqnarray}\label{#1} }
\newcommand{\eea}{\end{eqnarray}}

\newcommand{\refb}[1]{(\ref{#1})}
\newcommand{\p}{\partial}

\def\one{{\hbox{ 1\kern-.8mm l}}}
\def\zero{{\hbox{ 0\kern-1.5mm 0}}}

\begin{document}

\baselineskip 24pt

\begin{center}
{\Large \bf
$AdS_3/CFT_2$ to $AdS_2/CFT_1$}

\end{center}

\vskip .6cm
\medskip

\vspace*{4.0ex}

\baselineskip=18pt

\centerline{\large \rm   Rajesh Kumar Gupta and  Ashoke Sen}

\vspace*{4.0ex}

\centerline{\large \it Harish-Chandra Research Institute}

\centerline{\large \it  Chhatnag Road, Jhusi,
Allahabad 211019, INDIA}

\vspace*{1.0ex}
\centerline{E-mail:  rajesh@mri.ernet.in, 
sen@mri.ernet.in, ashokesen1999@gmail.com}

\vspace*{5.0ex}

\centerline{\bf Abstract} \bigskip

It has been suggested that the 
quantum generalization
of the Wald entropy for an extremal black hole is the logarithm of the
ground state degeneracy of a dual quantum mechanics in a fixed charge
sector. We test this proposal for supersymmetric
extremal BTZ black holes for which
there is an independent definition of the quantum entropy as the
logarithm of the degeneracy of appropriate states in the 
dual 1+1 dimensional
superconformal field theory. We find that the two proposals agree. This
analysis  also suggests a possible route to deriving the OSV
conjecture.

\vfill \eject

\baselineskip=18pt

Wald's formula for black hole 
entropy\cite{9307038,9312023,9403028,9502009}, 
when applied to extremal
black holes, leads to the entropy function 
formalism\cite{0506177,0606244}.
Since extremal black holes have an $AdS_2$ factor in their
near horizon geometry\cite{0705.4214,0803.2998}, one expects 
that the underlying quantum
gravity theory in this background will have a dual description
in terms of a conformal quantum mechanics (CQM) living at
the boundary of 
$AdS_2$\cite{9809027,9810251,9812073,9904143,
9910076,0008058,0009185,0412294,0710.2956,
0803.3621,0805.0095,0805.1861}.
In \cite{0805.0095} it was shown that in the classical limit, when
Wald's formula is expected to be valid, the Wald entropy
computed from the entropy function can be
interpreted as the logarithm of the ground state degeneracy of
this dual CQM in a fixed charge sector. 
This suggested that the latter should be taken as
the definition of the entropy of extremal black holes 
in the full quantum
theory.

In this paper we shall test this proposal for a special class of
black holes, -- the BTZ black holes\cite{9204099}. 
The latter are rotating
black hole solutions in $AdS_3$ characterized by their mass $M$
and angular momentum $J$. 
We shall assume that the BTZ black hole solution
has been embedded in a
string theory with certain amount of supersymmetry where we
have sufficient control on the system\cite{9711138,9712251}. 
In particular in this case via
$AdS_3/CFT_2$ 
correspondence\cite{9711200,9802109,9802150,9806087}
one can identify the BTZ black
holes as states in the  superconformal
field theory (CFT) 
living on the boundary of $AdS_3$, with the 
identification\footnote{$L_0$ and $\bar L_0$ denote the
Virasoro generators on the cylinder; thus in their definition 
we include the contributions $-c/24$ and $-\bar c/24$ of the central
charges.}
\be \label{eident}
L_0={M+J\over 2}, \qquad \bar L_0 = {M-J\over 2}\, .
\ee 
Extremal supersymmetric BTZ
black holes, corresponding to $M=\pm J$, correspond to states with
$\bar L_0=0$ and $L_0=0$ respectively. For definiteness 
we shall consider black holes with $M=J$, \i.e.\ with 
$\bar L_0=0$. In order that the state preserves supersymmetry it
must belong to the Ramond sector of the anti-holomorphic part of the
superconformal algebra of the CFT, so that the condition 
$\bar L_0=0$ forces the state to be in the 
supersymmetric ground
state of the Ramond sector\cite{9310194,9712251,0706.3359}. 

The identification of the BTZ black hole with a state in the dual CFT
suggests a natural definition of the entropy of this black hole, -- it is
simply the logarithm of the degeneracy of the corresponding states in
the CFT\cite{9712251}. 
For large $L_0$ where we can use 
Cardy formula to
estimate the degeneracy of states, the entropy defined this way agrees
with the one computed via Wald's 
formula\cite{9909061,0506176,0601228,0711.4671}. 
Our goal will be
to compare the definition of the quantum entropy of the black hole
based on the degeneracies in the dual CFT
with the one suggested by the $AdS_2/CFT_1$ correspondence,
where we identify the entropy as the logarithm of the degeneracy
of certain states in the dual CQM. Thus for this comparison we need to
study the relationship between the CQM and the CFT. The comparison
is not completely straightforward since the CFT lives 
on the boundary of the
$AdS_3$ space in which the black hole is embedded, whereas the
CQM lives on the boundary of $AdS_2$ that appears in the near
horizon geometry of the black hole.

The general BTZ black hole solution in an $AdS_3$ space
with scalar curvature $-6/l^2$ is given by
\be \label{e1}
 ds_3^2 = -{
(\rho^2 - \rho_+^2) (\rho^2 - \rho_-^2)\over l^2 \rho^2} d\tau^2
+ {l^2 \rho^2 \over (\rho^2 - \rho_+^2) (\rho^2 - \rho_-^2)} d\rho^2
+ \rho^2 \left(dy - {\rho_+ \rho_-\over l \rho^2} d\tau\right)^2\, ,
\ee
where $\tau$ denotes the time coordinate, $\rho$ is the radial variable,
$y$ is the azimuthal angle with period $2\pi$ and 
$\rho_\pm$ are parameters labelling the black hole solution
satisfying $\rho_+>\rho_-$.
$M$ and $J$ are determined in terms of $\rho_\pm$, but the
precise relation requires the knowledge of higher derivative terms.
Nevertheless 
the extremal limit always corresponds to $\rho_+\to \rho_-$.
Following \cite{0805.0095} we take this limit by 
first defining new variables
$\lambda$, $t$, $r$, $\phi$ and $R$ through
\be \label{e2}
\rho_+ - \rho_- = 2\lambda, \quad \rho - \rho_+ = \lambda (r-1),
\quad \tau = l^2 \, t / (4 \lambda), \quad y =
\phi + {l\over 4\lambda} \, \left( 1 - {2\lambda\over \rho_+}\right)
t , \quad \rho^+ = {lR\over 2}\, ,
\ee
and then
taking
$\lambda\to 0$ with $t$, $r$, $\phi$ and $R$ fixed.
In this limit the metric \refb{e1} takes the form
\be \label{e3}
ds_3^2 = {l^2 \over 4} \left[ -(r^2 - 1) dt^2 + {dr^2\over r^2-1}
+ R^2 \left(d\phi + {1\over R} (r-1) dt\right)^2 \right]\, .
\ee
The metric \refb{e3} is locally $AdS_3$. Thus
by the standard rules of AdS/CFT correspondence any quantum theory
of gravity in the background \refb{e3} has a dual (1+1) dimensional
conformal field theory.
Since locally this $AdS_3$ space is the same
as the one in which we embed the BTZ black hole, we expect that
as a local field theory the (1+1) dimensional CFT living on the 
boundary of the near horizon geometry of the BTZ black hole
must be identical to that living on the boundary of the $AdS_3$ in
which the full BTZ black hole solution is embedded. The conformal
structure of the two dimensional space in which the theory lives
will however be quite different for the theory dual to $AdS_3$ and the
one dual to the near horizon geometry of the black hole.

Now via 
a dimensional reduction we can also regard the three dimensional
metric \refb{e3} as a two dimensional field 
configuration\cite{9809027,9906078}. 
For this we
introduce a two dimensional metric $ds_2^2$, a scalar field $\chi$ and
a gauge field $a_\mu$ via the relation:
\be \label{e5}
ds_3^2 = ds_2^2 + \chi \, (d\phi + a_\mu dx^\mu)^2 \, ,
\ee
where $\{x^\mu\}$ for $\mu=0,1$ represent the two dimensional 
coordinates $(t,r)$.
{}From the two dimensional viewpoint, the background \refb{e3} 
takes the form
\be \label{e6}
ds_2^2 = {l^2 \over 4} \left[ -(r^2 - 1) dt^2 + {dr^2\over r^2-1}
\right], \quad \chi={l^2 \, R^2\over 4}, \quad a_\mu dx^\mu =
{1\over R} (r-1) dt\, .
\ee
\be \label{e7}
e\equiv F_{rt} = 1/R\, .
\ee
This describes an $AdS_2$ space-time with background scalar
and electric field. Then via the rules of AdS/CFT correspondence
the theory is dual to a CQM living on the
boundary of $AdS_2$. In particular 
we can relate the partition function of the quantum
gravity theory on $AdS_2$ to the partition function of the 
 CQM
living on the boundary of $AdS_2$\cite{0805.0095}.

Since \refb{e3} and \refb{e6} describe the same background,
the quantum theories dual to them must also be identical.
Consequently the CQM living
on the boundary of \refb{e6} and the (1+1)
dimensional CFT living on the boundary of \refb{e3} 
are also different
descriptions of the same quantum theory. Our goal will be to exploit
this equivalence to learn about the CQM living on
the boundary of $AdS_2$.

First consider the two dimensional viewpoint. The metric is that
of $AdS_2$, and the boundary is located at $r=r_0$. 
The induced metric, scalar and gauge field on the boundary are
\be \label{e7a}
ds_B^2 = - {l^2 \over 4}   (r_0^2 - 1) dt^2, \qquad
\chi_B = {l^2 R^2\over 4}, \qquad \left.
a_t\right|_B = {1\over R} (r_0-1)\, .
\ee
We shall denote by $H_t$
the total Hamiltonian of the CQM living on the
boundary of $AdS_2$ including the effect of the background
gauge fields and by $Q$ the conserved charge in
the CQM
conjugate
to the gauge field $a_\mu$ in 
the bulk.\footnote{In the analysis of 
\cite{0805.0095} the Hamiltonian was split into two parts, one
due to the background gauge fields given by $-a_t Q$
and the other due to the rest of the fields. We shall not need to
use this split.  Also the analysis
of \cite{0805.0095} was carried out using the rescaled time
coordinate $\wt t=r_0 t$
so that the metric on the boundary remains finite in the
$r_0\to\infty$ limit, but the span
of the time coordinate becomes infinite in this limit. This
corresponded to taking the infrared cut-off to infinity keeping the
ultraviolet cut-off fixed. In this paper we shall use the opposite
(and more conventional) viewpoint where we take $t$ as the
time coordinate. In this case the induced metric \refb{e7a} on the
boundary goes to infinity as $r_0\to\infty$ but the range of $t$
remains fixed. This corresponds to
taking the ultraviolet cut-off to 
zero keeping the infrared cut-off fixed.}

We now turn to the three dimensional viewpoint. The dual
(1+1) dimensional CFT lives
on the two dimensional 
boundary labelled by $(t,\phi)$ with induced metric
\be \label{ei1}
ds_B^2 = 
{l^2 \over 4} \left[ -(r_0^2 - 1) dt^2  
+ R^2 \left(d\phi + {1\over R} (r_0-1) dt\right)^2 \right]\, .
\ee
To get
some insight into this theory we introduce new coordinates
\be \label{ei2}
\wt t = R^{-1}\, \sqrt{r_0^2 - 1} \, t, \quad \wt\phi = 
\phi + {1\over R} (r_0-1) t\, ,
\ee
so that the metric \refb{ei1} becomes
\be \label{ei3}
ds_B^2 = {l^2 R^2 \over 4} [ -d\wt t^2 + d\wt\phi^2]\, .
\ee
Thus up to the overall scale factor the metric is the standard Minkowski
metric, and the space coordinate $\wt\phi$ is compact with period
$2\pi$. This gives a standard 1+1 dimensional CFT on a cylinder,
and the generators $i\p_{\wt t}$ and $-i\p_{\wt\phi}$ are
identified as\be \label{egen}
i\p_{\wt t} = L_0+\bar L_0, \qquad  -i\p_{\wt\phi}=
L_0-\bar L_0\, .
\ee
In order that in the extremal limit we get a
supersymmetric black hole, we impose Ramond boundary
condition along $\wt\phi$
on the anti-holomorphic part of the superconformal
algebra.

In relating this (1+1) dimensional CFT to the
CQM living on the boundary of $AdS_2$,
we must
identify 
the total Hamiltonian $H_t$ of the CQM as
the generator of $t$-translation in the CFT.
On the
other hand the charge $Q$ of the CQM can be
identified as the generator of $\phi$ translation.
This gives
\ben \label{ei4}
H_t &=& i\p_t = i R^{-1}\, \sqrt{r_0^2 - 1}{\p\over \p{\wt t}} + 
i {r_0-1\over R} {\p\over \p \wt\phi} = 2 \, R^{-1} r_0 \bar L_0 
+ R^{-1} (L_0-\bar L_0) + \OO(r_0^{-1}) \, , \nonumber \\
Q &=& -i \p_\phi = -i\p_{\wt\phi}=L_0 - \bar L_0 \, .
\een
Thus in the $r_0\to\infty$ limit, the only states with finite
$H_t$ eigenvalues are those with minimal value of $\bar L_0$. Since
we have Ramond boundary condition, the minimal value of $\bar L_0$
is 0. In other words the states
of the CQM living on the boundary of $AdS_2$ are
described by the $\bar L_0=0$ states of the 
1+1 dimensional CFT living
on the boundary of $AdS_3$.\footnote{This is in accordance with the
expectation that the  CQM dual to gravity in $AdS_2$
is described by the chiral half of the (1+1) dimensional CFT dual to
gravity in 
$AdS_3$\cite{9803231,9809027,9906078,0803.3621}.}
In particular the ground state degeneracy $d(q)$ of the CQM,
carrying a given charge $q$, can be identified as the degeneracy of
the states of the CFT which are in the ground state of the Ramond
sector in the anti-holomorphic sector and carries $(L_0-\bar L_0)$
eigenvalue $q$. The former is the quantity that appears in the
definition of the entropy via $AdS_2/CFT_1$ 
correspondence\cite{0805.0095}
whereas the latter appears in the definition of the entropy of
the extremal BTZ black hole via $AdS_3/CFT_2$ correspondence.
Thus we see that the two definitions of entropy agree up to
subtleties involving ultraviolet cut-off of the CFT to be discussed
below \refb{ecs11}

Using the identification of the CQM as a specific compactification
of the CFT we can compute the partition function of the theory.
For this we
make the Euclidean continuation $t\to -i u$. Regularity
of the metric \refb{e3} (or \refb{e6})
at the horizon $r=1$ requires $u$ to be a periodic
coordinate with period $2\pi$. {}From the point of view of the
CQM, the partition function
of the theory will be given by $Tr(e^{-2\pi H_t})$. Using \refb{ei4}
this can be reinterpreted as an appropriate trace over the Hilbert
space of the (1+1) dimensional CFT dual to gravity in $AdS_3$.
It is however instructive to do this computation directly in the
CFT. For this we note that under the replacement $t\to -i u$ the
boundary metric \refb{ei1} takes the form
\be \label{ej1}
ds_B^2 = 
{l^2 \over 4} \left[ (r_0^2 - 1) du^2  
+ R^2 \left(d\phi - {i\over R} (r_0-1) du\right)^2 \right]
= {l^2 R^2 \over 4} [ \tau_2^2 du^2 + (d\phi + \tau_1 du)^2]\, ,
\ee
where
\be \label{ej2}
\tau_1 = - {i\over R} (r_0-1), \quad \tau_2 = {\sqrt{r_0^2-1}\over R}\, .
\ee
The metric is complex, but we can nevertheless go ahead and
compute the partition function. Since $u$ and $\phi$ both have
period $2\pi$, the partition function of the CFT with this
background metric will be given by
\be \label{ej3}
Z = Tr\left[ e^{2\pi i (\tau_1 + i\tau_2) L_0 - 2\pi i (\tau_1 - i \tau_2)
\bar L_0}
\right] = Tr \left[ e^{-4\pi r_0 R^{-1} 
\bar L_0 - 2\pi R^{-1} (L_0-\bar L_0)}
+ \OO(r_0^{-1})
\right]\, .
\ee
This agrees with $Tr(e^{-2\pi H_t})$ with $H_t$ 
given in \refb{ei4}.
Eq.\refb{ej3} again shows that in the $r_0\to\infty$ limit only the
$\bar L_0=0$ states contribute to the trace. We also see that in this limit
the contribution to the partition function from states with a given
charge $Q=q$ is given by 
\be \label{ej4}
d(q) \, e^{ -2\pi e q}\, ,
\ee
where $q$ is the $L_0-\bar L_0$ eigenvalue, $e=1/R$ is the
near horizon electric field, and $d(q)$ is the degeneracy of
the states with charge $q$. Eq.\refb{ej4} agrees with eq.(24)
of \cite{0805.0095}, where this result was also derived both
from the microscopic computation in the CQM and
a computation of the partition function in the bulk theory in the
semiclassical limit.

A similar dimensional reduction from $AdS_3$ to $AdS_2$
was carried out in \cite{0608021} in the context of extremal
black holes in type IIA
string theory on a Calabi-Yau manifold. However in that paper the
authors interpreted the $\phi$ coordinate as the euclidean time
direction and the $u$ coordinate as the spatial circle, thereby
arriving at a modular transformed version of eq.\refb{ej3}.
Since our goal is to identify the CQM living on the boundary of
$AdS_2$, we must choose $u$ as the time coordinate on
the boundary of $AdS_3$ so that it
matches the
time coordinate of the CQM.

So far in our analysis we have considered neutral BTZ black
holes. Let us now suppose that 
the three dimensional theory has additional $U(1)$ gauge
fields $A^{(i)}_M$ with Chern-Simons action of the form
\be \label{ecs1}
{1\over 2} \int d^3 x \, \epsilon^{MNP} \, C_{ij}\,
A^{(i)}_M 
F^{(j)}_{NP}, \qquad F^{(i)}_{NP}\equiv
\p_N A^{(i)}_P - \p_P A^{(i)}_N\, ,
\ee
where $M,N,P$ run over the three coordinates of $AdS_3$
and $C_{ij}$ are constants. 
Then we can construct charged black hole solutions 
by superimposing
on  the original BTZ solution \refb{e1} constant
gauge fields:
\be \label{ecs2}
A^{(i)}_M dx^M = w_i \left[
dy - {1\over l}\, {\rho_- \over \rho_+} \, d\tau\right]\, .
\ee
Here $w_i$ are constants. The term proportional to $d\tau$ has
been chosen so as to make the gauge fields non-singular at the
horizon. Even though the gauge field strength
vanishes, the background \refb{ecs2} induces a charge on the
black hole since the latter, being proportional to $\delta S / 
\delta F^{(i)}_{\rho t}$ (in the classical limit), 
is given by $C_{ij} A^{(j)}_y$ up to
a constant of proportionality.  Taking the near horizon limit
as in \refb{e2} we arrive at the background
\be \label{ecs3}
ds_3^2 = {l^2 \over 4} \left[ -(r^2 - 1) dt^2 + {dr^2\over r^2-1}
+ R^2 \left(d\phi + {1\over R} (r-1) dt\right)^2 \right]\, ,
\qquad A^{(i)}_M dx^M = w_i d\phi\, .
\ee
In order to make contact with the two dimensional viewpoint
we define two dimensional gauge fields $a_\mu^{(i)}$ and scalar
fields $\chi^{(i)}$ via the relations:
\be \label{ecs4}
A^{(i)}_M dx^M = \chi^{(i)} (d\phi+ a_\mu dx^\mu) + 
a^{(i)}_\mu\, dx^\mu\, ,
\ee
where 
$a_\mu$ has been defined in \refb{e5}. For the background
\refb{ecs3} we have $a_\mu dx^\mu = {1\over R} (r-1) dt$,
and hence\cite{0708.1270}
\be \label{ecs5}
\chi^{(i)}=w_i, \quad a^{(i)}_\mu dx^\mu = e^{(i)} (r-1) dt, \quad
e^{(i)} \equiv - {w_i\over R}\, .
\ee
$e^{(i)}$ is the near horizon electric field associated with the two
dimensional gauge fields $a^{(i)}_\mu$. 

We shall now compute the partition function of the CQM
living on the boundary of $AdS_2$ in the presence of these
background gauge fields. This is equivalent to computing the
partition function of the CFT living on the boundary of the
space-time given in \refb{ecs3}. Let 
$(J_{(i)}^\phi,J_{(i)}^t)$
be the currents in the CFT dual to the gauge fields $A^{(i)}_M$
in the bulk. Then in the presence of the gauge field background
given in \refb{ecs3} we have an insertion of
\be \label{ecs6}
\exp\left[ i w_i \int dt d\phi \sqrt{-\det g} J_{(i)}^\phi\right]\, ,
\ee
in the boundary theory.
To proceed further we
need to assume some properties of the currents $J_{(i)}$.
Typically in $AdS_3/CFT_2$ correspondence the currents dual
to gauge fields are either holomorphic or anti-holomorphic
depending on the sign of the Chern-Simons term in the bulk
theory\cite{0609074}. We shall assume for simplicity
that all our gauge fields
are dual to holomorphic currents; if the state carries charge
associated with 
anti-holomorphic currents then in general we shall not be
able to satisfy the $\bar L_0=0$ condition and the analysis
will be more complicated.\footnote{If there are
gauge fields dual to  anti-holomorphic currents, then an analysis
identical to that for the holomorphic currents shows that
in the first term in the exponent in eq.\refb{ecs10}, $\bar L_0$
will be replaced by
$\bar L_0+\sum'_i w_i Q_{(i)}$, 
with the sum over $i$ in
$\sum'$ running over the anti-holomorphic currents. 
The finite part retains 
the same form as the
holomorphic currents, \i.e.\
$-2\pi \sum' e^{(i)} Q_{(i)}$, in agreement with the results
of \cite{0805.0095}.}
This gives a
relation between $J_{(i)}^\phi$ and $J_{(i)}^t$. To determine
this relation we note from \refb{ej1} that in the euclidean theory
the holomorphic
coordinate $z$ is given by $\phi + \tau_1 u + i \tau_2 u$. Using
the relation $u=it$ and the values of $\tau_1$, $\tau_2$ given in
\refb{ej2} we get
\be \label{ecs6a}
z = \phi - {1\over R} t + \OO(r_0^{-1})\, .
\ee
Requiring holomorphicity gives $J_{(i)}^z=0$ since by virtue of
current conservation $\p_z J_{(i)}^z=0$, $J_{(i)}^z$ would
have described an anti-holomorphic current. Thus we have
\be \label{ecs7}
J_{(i)}^\phi - {1\over R} J^t_{(i)}=0\, .
\ee
Substituting this into \refb{ecs6} and using the definition of
the charge $Q_{(i)}$,
\be \label{ecs8}
Q_{(i)} = \int   d\phi \sqrt{-\det g} J_{(i)}^t\, ,
\ee
we can express \refb{ecs6} as
\be \label{ecs9}
\exp\left[ i w_i \int dt\,  Q_{(i)}/R \right] = 
\exp(2\pi\, w_i \, Q_{(i)}/R) = \exp( - 2\pi \, e^{(i)} Q_{(i)})
\, ,
\ee
where in the last step we have used \refb{ecs5}. Insering this
into \refb{ej3} and using $e=1/R$ we get
\be \label{ecs10}
Z = Tr \left[ e^{-4\pi r_0 R^{-1} \bar L_0 - 2\pi \sum_I e^I Q_I}
\right]\, ,
\ee
where the index $I$ now runs over all the two dimensional
gauge fields, -- the one coming from the dimensional reduction
of the three dimensional metric as well as the ones coming from the
three dimensional gauge fields. {}From \refb{ecs10} we see that
in the $r_0\to\infty$ limit we are still restricted to the $\bar L_0=0$
states. The contribution from the sector
with charge $\vec q$ is given by
\be \label{ecs11}
d(\vec q) \, e^{-2\pi \sum_I q_I e^I}\, ,
\ee
in agreement with eq.(24) of \cite{0805.0095}. 
Here $d(\vec q)$
denotes the degeneracy of $\bar L_0=0$ states in the CFT carrying 
charge $\vec q$. It can also be interpreted as the 
degeneracy of the
lowest energy states in the CQM carrying charge $\vec q$.

One issue that
we have not completely resolved is the following. From
\refb{ei3} we see that in the $(\wt t, \wt \phi)$ coordinate
system the conformal factor in front of the metric remains
finite as $r_0\to\infty$, suggesting that we have a finite 
ultraviolet cut-off. In particular the size of the $\wt\phi$ circle
is of the order of the cut-off. We do not have a direct understanding
of the role of this cut-off in the CFT. However studying the effect of
this cut-off in the bulk gives us some insight. First of all note that in
conventional $AdS_3$, it is more natural to define the partition
function by summing over states of all charges with a fixed value
of the chemical potential. However in $AdS_2$ the modes representing
fluctuation of the total charge represent non-normalizable deformations
and hence it is more natural to define the partition function by
summing over a fixed charge sector\cite{0805.0095}. 
Thus it would seem that the
effect of the finite ultraviolet cut-off in the CFT must be to
restrict the Hilbert space of a given CFT to a fixed charge sector.
There are also other effects of this finite cut-off in the bulk when
we embed the BTZ black hole in a supersymmetric 
theory with additional moduli
scalars and vector fields.
When we view the extremal BTZ black
hole from the point of view of the asymptotically $AdS_3$ space-time
by setting $\rho_+=\rho_-$ in \refb{e1} 
then the ultraviolet cut-off is small compared to the size of the $y$
circle since the latter approaches $\infty$ as $\rho\to\infty$, 
but such asymptotic space-time could admit other multi-centered
black hole solutions\cite{0802.2257}. On the other hand 
when we view the same extremal
black hole from the point of view of its near horizon geometry as in
\refb{e3},  then the size of the $\phi$ 
circle becomes comparable to the
ultra-violet cut-off, but this space-time 
geometry no longer admits the
other multi-centered black hole solutions
in $AdS_2$ since the values of the various scalar
fields are fixed at their attractor 
values.\footnote{Possible exceptions are multi-centered black holes
with mutually local charges\cite{9202037,9812073,0504221},
\i.e.\ charges satisfying 
$(\vec q_i\cdot \vec p_j-\vec q_j\cdot \vec p_i)=0$ 
where $(\vec q_i, \vec p_i)$ denote the electric and magnetic charge
vectors of the $i$th black hole.
But they do not contribute to
the degeneracy\cite{0206072,0702146}.}
Thus it would seem that the ultraviolet cut-off
weeds out the contribution due to the
multi-centered black hole configurations of the type discussed in
\cite{0802.2257} 
from the CFT spectrum. 
In support of this
speculation we would like to note that for large $R$ the 
size of the $\phi$ circle is large compared to the
ultra-violet cut-off and hence effect of the
cut-off is expected to be small. This is precisely the region in
which the entropy of a single centered black hole gives the
dominant contribution to the entropy\cite{0802.2257}.

Even though it is more natural to work in a fixed charge sector
of $AdS_2$,
one can get some insight into the OSV conjecture
if one does sum over the contribution from different charge sectors. 
After summing over charges the full partition function is given by
\be \label{ej5}
Z(\vec e) = \sum_{\vec q} d(\vec q) \, e^{-2\pi \vec e\cdot \vec q}\, .
\ee
For large charges the dominant contribution to this sum comes
from $\vec q$ satisfying $\p\ln d(\vec q) / \p q_I = 2\pi e^I$, in
agreement with the classical relation between the electric field and
the charge.
The right hand side of \refb{ej5}
has the flavor of the black hole partition function defined
in \cite{0405146}. 
On the other hand, using $AdS/CFT$
correspondence, the left hand side 
can be expressed as a functional integral
over the fields in the bulk 
theory.\footnote{Note that we have switched back from
the three dimensional viewpoint to the two dimensional 
viewpoint. The black hole
partition function has been analyzed using 
AdS/CFT correspondence
earlier (see {\it e.g.} \cite{0005003,0607138,0608059}). 
Also various other approaches to relating the entropy function
formalism to Euclidean action formalism and / or 
OSV conjecture can be found in 
\cite{0605279,0704.0955,0704.1405}.
The advantage of 
our approach lies in the fact that
since we apply $AdS/CFT$ correspondence on the near
horizon geometry, the chemical potentials dual to the charges
are directly related to the near horizon electric field, and hence,
via the attractor mechanism, to other near horizon field
configuration. 
Furthermore the path integral needs to be performed only
over the near horizon geometry where we have enhanced 
supersymmetry and hence stronger non-renormalization
properties. The approach closest to ours is the one given in
\cite{0608021}; we shall comment on it later.
A different approach to deriving the OSV conjecture
using AdS/CFT correspondence can be found in \cite{0602046}.}
Now, as was shown in \cite{0805.0095}, 
after ignoring terms linear in $r_0$ in the
exponent -- which must cancel among themselves --
the classical result for the partition
function in the $r_0\to\infty$ limit is given by
\be \label{ej66}
Z = e^{-2\pi f}\, ,
\ee
where $f$ is the classical Lagrangian density evaluated in the
near horizon geometry. 
One might expect that the effect of
quantum corrections would be to replace the classical Lagrangian
density by some effective Lagrangian density. 
As we shall now review, if we assume that the effective
Lagrangian density that contributes to the partition function
is governed 
only by the $F$-type
terms,  \i.e.\ terms which
can be encoded in the prepotential
$\FF$\cite{9602060}, then $Z$ takes the form
predicted in the original OSV conjecture.

\def\FF{F}

In $\NN=2$ supergravity theories in four dimensions
the information about the
`F-type terms' can be encoded in a function
$\FF(\{ X^I\}, \wh A)$ -- known as the prepotential --
of a set of complex variables $X^I$
which are in one to one correspondence with the gauge fields and
an auxiliary complex variable $\wh A$ related to the square of
the graviphoton field strength\cite{9602060,9603191}. 
Supersymmetry demands that
$\FF$ is a homogeneous function of degree two in its
arguments:
\be \label{efcon}
\FF(\{\lambda X^I\}, \lambda^2 \wh A) = \lambda^2
\FF(\{ X^I\}, \wh A)\, .
\ee
For a given choice of electric field one finds that
the 
extremum of the effective Lagrangian density computed with the
$F$-term effective action occurs at the 
attractor point
where\cite{9508072,9602111,9602136,
9711053,9801081,9812082,9904005,9910179,
0007195,0009234,0012232,0603149}
\be \label{eatt}
\wh A=-4w^2, \quad 
4(\bar w^{-1} \bar X^I + w^{-1}  X^I) = e^I, \quad
4(\bar w^{-1} \bar X^I - w^{-1}  X^I) = - i p^I\, .
\ee
Here $w$ is an arbitrary complex parameter and $p^I$ are the
magnetic charges carried by the black hole. These magnetic
charges have not
appeared explicitly in our discussion so far because from the
point of view of the near horizon geometry they represent
fluxes through compact two cycles and
appear as parameters labelling
the two (or three) 
dimensional field theory describing the near horizon
dynamics.
The value of the effective Lagrangian density at the
extremum \refb{eatt} is given by\cite{0603149}
\be \label{eatt2}
f = 16 \, i \, (w^{-2}\FF - \bar w^{-2}\bar \FF)\, .
\ee
Note that \refb{eatt} determines $X^I$ in terms of the
unknown parameter $w$. However due to the scaling symmetry
\refb{efcon}, $f$
given in \refb{eatt2} is independent of $w$.
Using
this scaling symmetry we can choose
\be \label{eatt3}
w = - 8 \, i\, ,
\ee
and rewrite \refb{eatt}, \refb{eatt2} as
\be \label{eatt4}
\wh A=256, \quad X^I = - i (e^I + i p^I)\, ,
\ee
\be \label{eatt5}
f= -{i\over 4} 
(\FF(\{X^I\}, 256) - \overline{\FF(\{X^I\}, 256)})
\, .
\ee
Thus we have
\be \label{ezexp}
Z(\vec e) = e^{-\pi \, Im \, \FF(\{p^I - i e^I\}, 256)}
\, .
\ee
This is precisely the original OSV conjecture\cite{0405146}.

It has however been suggested in subsequent papers that agreement
with statistical entropy requires modifying this formula by including
additional measure factors on the right hand side
of \refb{ezexp}\cite{0508174,0601108,0702146}. A careful analysis
of the path integral keeping track of the holomorphic 
anomaly\cite{9302103,9307158,9309140}
may be able to
reproduce these corrections, but we shall not undertake that task
here.
Some of these corrections are in fact
necessary for restoring
the duality invariance of the final result for the entropy\cite{0601108}.

Ref.\cite{0608021} presented an argument as to why the
partition function of type IIA string theory on $AdS_2\times S^2
\times CY_3$ may be related to $|Z_{top}|^2$. 
In this analysis the divergence due to the integration over $AdS_2$
was regulated by supersymmetry.
This argument led
to $Z_{AdS_2}= |Z_{top}|^{2C}$, where $C$ is a constant that
was not calculated directly from first principles. 
In our interpretation of the $AdS_2$ partition function there is a
clear understanding of the divergent parts that is independent
of supersymmetry, -- terms linear in $r_0$
in the exponent represent the effect of ground state energy and the
$r_0$ independent piece encodes information about the
ground state spectrum. 
In particular the classical partition function calculated 
with F-type terms in
our approach agrees
with $|Z_{top}|^2$ {\it after
we remove the terms linear in $r_0$ from the exponent.}
Thus combining this regularization scheme with the
analysis of \cite{0608021} may lead to a complete
understanding of $Z_{AdS_2}$. In particular there may be
additional finite pieces from the interference between order
$r_0$ divergent terms and order $r_0^{-1}$ terms
which reproduce the measure
factors described in \cite{0508174,0601108,0702146}.

Our attempt to justify the OSV conjecture from
a macroscopic viewpoint makes it clear that 
$d(\vec q)$ appearing in the expression for the black hole
partition function counts only the states associated with
single centered black holes.\footnote{An operational 
definition of such a $d(\vec q)$ 
can be taken, for example, as the
degeneracy of microstates evaluated at the attractor point
corresponding to $\vec q$.}
Thus OSV formula should
have nothing to say about the contribution to the entropy
from the
multi-centered black holes. This in particular would explain why we
do not see the effect of wall crossing or the entropy enigma
discussed in \cite{0702146} in the OSV formula.

{\bf Acknowledgement:} We would like to thank 
Andrew Strominger and Sandip Trivedi
for useful discussions.

\end{document}